\documentclass[twocolumn,showpacs,prl,superscriptaddress]{revtex4}
\usepackage{graphicx}
\usepackage{dcolumn}
\usepackage{bm}
\usepackage{amsmath}
\usepackage{ulem}
\def\ffi{\varphi}
\def\bL{{\bf L}}
\def\bk{{\bf k}}
\def\bK{{\bf K}}
\def\bq{{\bf q}}
\begin{document}

\title{Metal-insulator transition in the Hartree-Fock phase diagram of the
fully polarized homogeneous electron gas in two dimensions}
\author{B. Bernu}
\affiliation{LPTMC, UMR 7600 of CNRS, Universit\'e P. et M. Curie, Paris, France}
\author{F. Delyon}
\affiliation{CPHT, UMR 7644 of CNRS, \'Ecole Polytechnique, Palaiseau, France}
\author{M. Duneau}
\affiliation{CPHT, UMR 7644 of CNRS, \'Ecole Polytechnique, Palaiseau, France}
\author{M. Holzmann}
\affiliation{LPTMC, UMR 7600 of CNRS, Universit\'e P. et M. Curie, Paris, France}
\affiliation{LPMMC, UMR 5493 of CNRS, Universit\'e J. Fourier, Grenoble, France}

\date{\today}
\begin{abstract}
We determine the ground state of the two-dimensional, fully polarized electron gas within the
Hartree-Fock approximation without imposing any particular symmetries on the solutions.
At low electronic densities, the Wigner crystal solution is stable, but
for higher densities ($r_s$ less than $\sim  2.7$)
 we obtain a ground state
of different symmetry: the charge density forms a triangular lattice with about $11\%$ more
sites than electrons.  We argue that 
this conducting state with broken
translational symmetry
remains the ground state of the high density region in the thermodynamic limit 
giving rise to a metal to insulator transition.  
\end{abstract}
\maketitle

The two-dimensional homogeneous electron gas is one of the fundamental models
in condensed matter physics. Despite its simplicity - the system consists
of electrons interacting through a $1/r$-potential to which a uniform
positive background is added for charge neutrality  - the phase diagram at zero temperature  is 
nontrivial\cite{Tanatar,Bernu,Waintal}. In general, it is given in terms of 
the dimensionless parameter $r_s=1/\sqrt{\pi na_B^2}$, 
where $n$ is the electronic density and $a_B$ the Bohr radius. 
At low density (large $r_s$), the potential energy dominates over the kinetic energy and the system
forms a perfect triangular lattice, the Wigner crystal (WC), whereas in the high density region 
($r_s \to 0$) the kinetic energy favors a uniform Fermi gas (FG) phase\cite{Tanatar}.
Already Wigner\cite{Wigner}  argued that the unpolarized FG is unstable even in the limit $r_s \to 0$.
Later, Overhauser showed the instability of the unpolarized WC with respect to spin-density waves, even within the Hartree-Fock approximation (HF)\cite{Overhauser}.
It has further been  conjectured that the Coulomb potential prevents
any  first order transition between the WC and a FG\cite{Spivac}. 
Despite these rather general
instability theorems, we are not aware of any quantitative calculations establishing the
true ground state of the electron gas within HF\cite{Giulani}.
A previous HF study of the two and three dimensional electron gas 
with imposed symmetries considered only the FG and different crystal structures for the WC phase\cite{Needs}, and only recently an unrestricted  HF study of the unpolarized three-dimensional electron gas was performed which proposes a more complicated structure of a ground state with spin-density waves in the high density region\cite{Shiwei}.
Indeed, establishing  the precise HF phase diagram of the electron gas is a fundamental question,  since many-body correlation effects can only be quantified with respect to the best HF solution.

In this letter, we consider the two-dimensional electron gas, and, for simplification,
we concentrate on the HF ground state of the completely
polarized system.
At low densities, $r_s \gtrsim 2.7$, our simulations always lead to a WC, but for higher densities
and large enough number of electrons, $N$, the solution is neither a FG nor a WC: the density modulation corresponds to a partially occupied crystal 
of different symmetry compared to the insulating WC phase
(the number of sites is larger than $N$). We refer to this solution as a metallic phase.

To observe the metallic phase, the number of electrons has to exceed a threshold 
ranging from 
$N \gtrsim 10$  at $r_s\sim2.7$ up to $N \gtrsim 10^2$ at $r_s\sim 1$.
For sizes below the threshold, the solution is the FG \cite{mphase}.
However we can  prove analytically that the metallic phase has a lower energy than the FG down to $r_s=0$ in the thermodynamic limit. \cite{Enerrs0}
This proof also indicates that the threshold size varies with $r_s$  as $\exp(Cst/r_s)$, which may explain why this metallic phase has not been observed in previous simulations.
It is possible to understand heuristically, why the metallic phase occurs  at small $r_s$. One expects that the electrostatic energy is always lowered by a periodic triangular charge density. On the other hand, at small $r_s$ the kinetic energy dominates and favors states close to the Fermi surface. The simplest solution is obtained  by choosing a triangular  modulation lattice with
generators $Q_i$ of modulus $2k_F$. This modulation is mainly carried by the wave vectors close to the Fermi surface.
These new HF solutions open a new perspective for the qualitative understanding of the experimental observed  metal to insulator transition\cite{Kravchenko94} and should be considered in studies beyond the HF approximation.

The N-body Hamiltonian, $H=K+V$, contains the kinetic energy $K$
and the $1/r$-periodic Coulomb potential $V$ where a uniform positive charge background is 
subtracted. 
Within the HF approximation, the search of the true ground state of the
quantum many-body system
is reduced to the simpler problem of finding the lowest energy states in the subset of the Slater determinants. Let $\Phi=\ffi_1\wedge\cdots \wedge\ffi_N$ be the Slater determinant associated with the single particle states $\{\ffi_i\}$
and $E(\ffi_1,\ffi_2,\dots,\ffi_N)$ the corresponding energy expectation value.
The variation of $E$ with respect to a variation 
of the single particle state
$\delta\ffi_i$
is then given by
\begin{eqnarray}
\delta E&=&\sum_i \left<h_\Phi\ffi_i | \delta\ffi_i\right>+
\sum_i \left<\delta\ffi_i | h_\Phi\ffi_i\right>.
\end{eqnarray}
where $h_\Phi$, the so-called HF Hamiltonian, is a single particle operator depending on the full state $\Phi$ (not on the particular choice of the $\ffi_i$'s). Extremal states must satisfy the following
equation
 \begin{align}
\label{EqH}
h_\Phi\ffi_i=\sum_j C_{ij}\ffi_j,
\end{align}
where $C_{ij}$ are the Lagrange coefficients associated with the normalization constraint 
$\left<\ffi_i|\ffi_j\right>=\delta_{ij}$. Conversely, if $\Phi=\ffi_1\wedge\cdots \wedge\ffi_N $ is not an extremum, we have
 \begin{align}
\label{EqH1}
h_\Phi\ffi_i=\sum_j C_{ij}\ffi_j+\theta_i
\end{align}
where the $\theta_i$'s satisfy $\left<\ffi_i|\theta_j\right>=0$, $\forall i,j$.
Within the steepest descent method
one chooses first a $N\times N$ unitary transformation $A=(a_{ij})$ such that one obtains
\begin{align}
\left<\ffi'_i| \theta'_j\right>=0, \quad 
\left<\theta'_i | \theta'_j\right> \propto \left<\ffi'_i | \ffi'_j\right>=\delta_{ij},
\quad  \forall i,j
\end{align}
for the transformed single particle states
$\theta'_i=\sum_ja_{ij}\theta_j$, $\ffi_i'=\sum_ja_{ij}\ffi_j$.
The energy $E(\ffi_1+\lambda\theta_1,...,\ffi_N+\lambda\theta_N)$ can be expressed as a sum of rational fractions whose numerators and denominators are polynomials of order four, at most. Thus, it is possible to find the best $\lambda$ and to iterate the process until a stationary state is reached.

In fact, this method has the same drawbacks as the steepest decent method in linear 
optimization problems; in general, it converges slowly. For linear problems, 
conjugate gradient methods are preferable \cite{CG, FDMD}. 
However, since the HF states do not form  a linear space, the genuine conjugate gradient method does
not apply here. We have therefore adapted a variant of this method to the non-linear case.
Let $\eta_i$ be the previous variation $\delta\ffi_i$,  and $\theta_i$ is obtained 
by Eq.\,(\ref{EqH1}).
We then  compute $E(\ffi_1+\lambda\theta_1+\mu\eta_1,\cdots,\ffi_n+\lambda\theta_N+\mu\eta_N)$ for
six values of the pairs $\{\lambda,\mu\}$ in order to  approximate $E$ by a polynomial of order two in 
$\lambda$ and $\mu$. 
Minimizing the polynomial with respect to $\lambda$ and $\mu$,
we obtain the new changes of the single particle states, $\delta\ffi_i$, and the corresponding
energy change. This process is iterated
until the relative variation of the energy, $\delta E/E$, is sufficiently small.

\begin{figure}
\begin{center}
\includegraphics[width=0.45\textwidth]{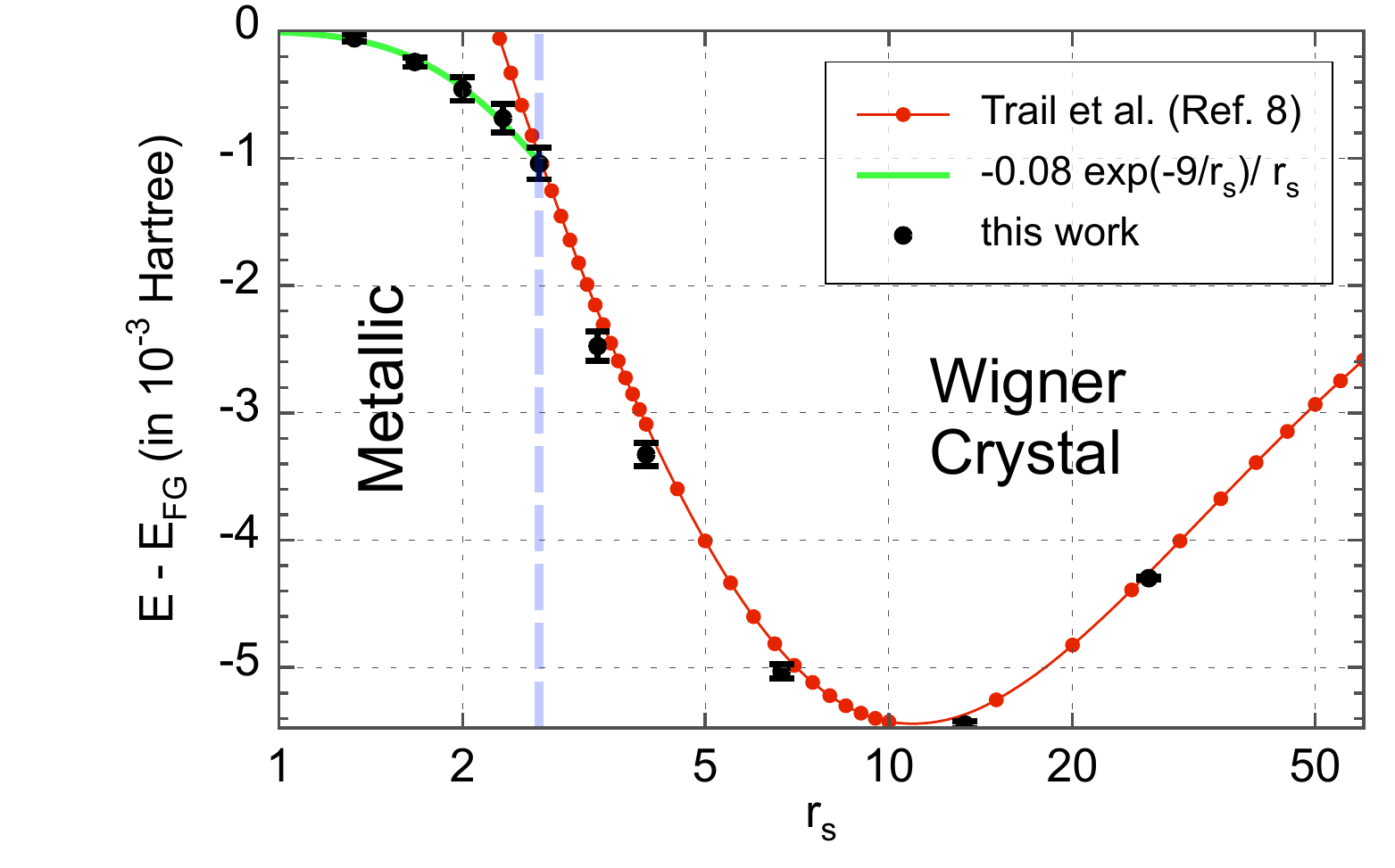}
\caption{Extrapolated energies $E_{\infty} - E_{FG}$ (milli-Hartree units) versus $r_s$. Points with error bars:  present calculations, full line (red): data of Ref.~\cite{Needs}, full line (green): fit to present results for ($r_s\lesssim 2.7$), vertical dash line (blue): $r_s\sim2.7$.
}
\label{FIG-Comparison}
\end{center}
\end{figure}

Since we expect the electrons to 
crystallize on a triangular lattice at  low densities,
we have chosen periodic conditions compatible with this geometry.
Thus, the unit cell  of our periodized system is given in terms of  two 
vectors $\left\{\bL_1,\bL_2\right\}$ of length $L$ 
and with an angle of 60 degrees between both; the  volume of the unit cell is $\Omega= L^2\sqrt{3}/2$. We have restricted our study to system sizes which are
compatible  both with the triangular lattice and with a closed shell occupation in $k$-space.
Any triangular crystal with unit-cell vectors
 $\left\{{\bf e}_1,{\bf e}_2\right\}$ is
 compatible with the boundary conditions if it satisfies $\bL_1=l {\bf e}_1+m{\bf e}_2$ and $\bL_2=-m {\bf e}_1+(l+m){\bf e}_2$, where $(l,m)$ are two non-negative integers. The
 number of sites of the lattice is given by $N_c=\det({\bf L}_1,{\bf L}_2)/\det({\bf e}_1,{\bf e}_2)=l^2+m^2+lm$. For a commensurate lattice $N=N_c$.
 
 We compute the wavefunction on a $N_g\times N_g$ grid, the fast Fourier transform is used 
 to switch between real and reciprocal space\cite{K-Wigner}.
We have systematically checked  the convergence of the solution  with respect to the grid size.
For the FG ground state, convergence is reached once all $k$-vectors up to $2k_F$ are represented in the grid ($N_g\sim 4\sqrt{N/\pi}$).
At larger $r_s$,  in the WC phase,
the wave functions are essentially gaussians\cite{VAR-GAUSSIAN}.
The width $\delta$ of the gaussians scales as $\delta/L \propto (r_s N)^{-1/2}$.  
For a correct resolution of the gaussians we need $L/N_g \propto \delta$, so that the
number of grid points increases  at low densities,  $N_g \propto (N r_s)^{1/2}$.
Convergence was reached 
for $N_g=32$ (resp. 64, 128) for $N\le43$ (resp. $N\le200$, $N\le500$) up to $r_s=30$. 
Whenever the number of grid points was chosen too small, solutions without any particular symmetries
have been obtained.

We have further studied the influence of the initial state on the final solution, choosing different types of wavefunction for  initialization:
a WC state, a converged state stored at larger or lower $r_s$, a  state
initialized with random numbers, or a ``metallic state'' as described below (this state has $N_c>N$ maxima in the charge density).


Typically, during the minimization procedure,
the energies decrease exponentially; 
however,
the rate of convergence depends on the initial state. The decrease in energy
during transitions to a different symmetry is in general much smaller than the convergence
within the same symmetry.
We have often seen energy plateaus with changes of relative energy $\lesssim 10^{-4}$ just
before the occurrence of a transition to a completely different state.
For system sizes up to $N=151$, the minimization is continued until a relative precision of $10^{-12}$ is reached, and for larger $N$ a relative precision of $10^{-5}$ is used.

We have studied systems with up to $500$ electrons at densities 
corresponding to $r_s=1$ up to $r_s=30$.
In Figure \ref{FIG-Comparison}, 
we report the energies of the obtained HF ground state $E_{\infty}(r_s)$,
extrapolated to the thermodynamic limit, as a function of $r_s$. 
In the high density region, for $r_s>3$, we obtain good agreement with the results of
Trail {\it et al.}\cite{Needs}, which imposed the triangular WC symmetry.
For $r_s<3$, we find ground states of different symmetries than the WC which we describe below. These new states have been eventually found down to $r_s=1$, 
at smaller $r_s$, the FG solution is stable for our finite system sizes with $N \le 500$. 
\begin{figure}
\begin{center}
\includegraphics[width=.35\textwidth]{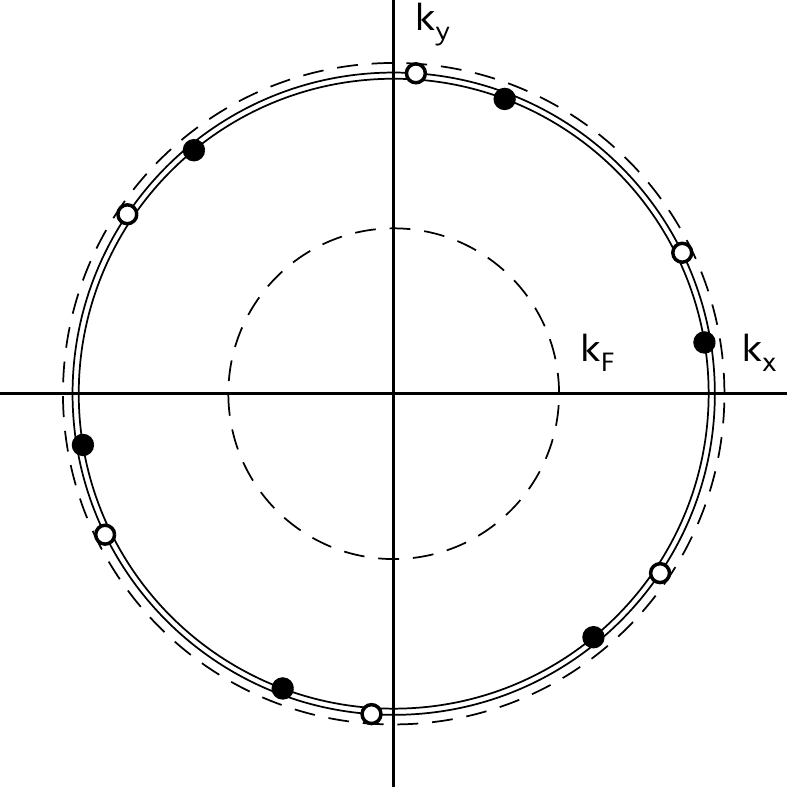}
\caption{Main points of $\tau(\bq)$ Eq.\,(\ref{EQ-PHIPERP}) for N=151. The black  (resp. white) dots corresponds to  $r_s=3.33$ (resp. 2.7). The radii of the dashed circles are $k_F$ and $2k_F$.
The ratio of the radii of the solid circles is $\sqrt{157/151}$. The other points in $\tau(\bq)$ are several order of magnitude less.
}
\label{FIG-allrs}
\end{center}
\end{figure}

We now turn to a more detailed analysis of the different phases.
It is instructive to follow the evolution of the Slater determinants in $k$-space as a function
 of $r_s$.
In order to quantify the differences to the FG solution,
 we project  each plane wave $ \phi_{\bk_0}$ of the FG ($k_0 \le k_F$), on our HF solution $\{ \ffi_i\}$,
\begin{eqnarray}
 \phi_{\bk_0}& =& \phi_{\bk_0}^\parallel+\phi_{\bk_0}^\perp,\\
 \phi_{\bk_0}^\parallel &=& \sum_i (\ffi_i,\phi_{\bk_0}) \ffi_i
\end{eqnarray}
 where $\phi_{\bk_0}^\parallel$ is the projection of $ \phi_{\bk_0}$ on the subspace spanned
 by the $\ffi_i$'s. The deviation of the HF solution from the 
 FG can be measured by $\tau( \bq )$ as
\begin{eqnarray}
\label{EQ-PHIPERP}
\tau( \bq ) &=& \sum_{k_0 \le k_F,k_0\neq q}  \left|\phi_{\bk_0}^\perp(\bq-\bk_0) \right|^2
\end{eqnarray}
For the FG, $\tau(\bq)$ is 0. Remarkably, for the WC or the metallic ground states, $\tau(\bq)$ is non zero only for few $\bq$-vectors  (see Fig. \ref{FIG-allrs}).
For the WC, $\tau(\bq)$ is nonzero on the points corresponding to the triangular reciprocal lattice: $a\bq_1+b\bq_2$, with $a$, $b$ integers and $\bq_1=l\bK_1-m\bK_2$, $\bq_2=m\bK_1+(l+m)\bK_2$ ($\bK_\alpha \cdot \bL_\beta=2\pi\delta_{\alpha\beta}$) \cite{LatticeModulus}.
\begin{figure}
\begin{center}
\includegraphics[width=.49\textwidth]{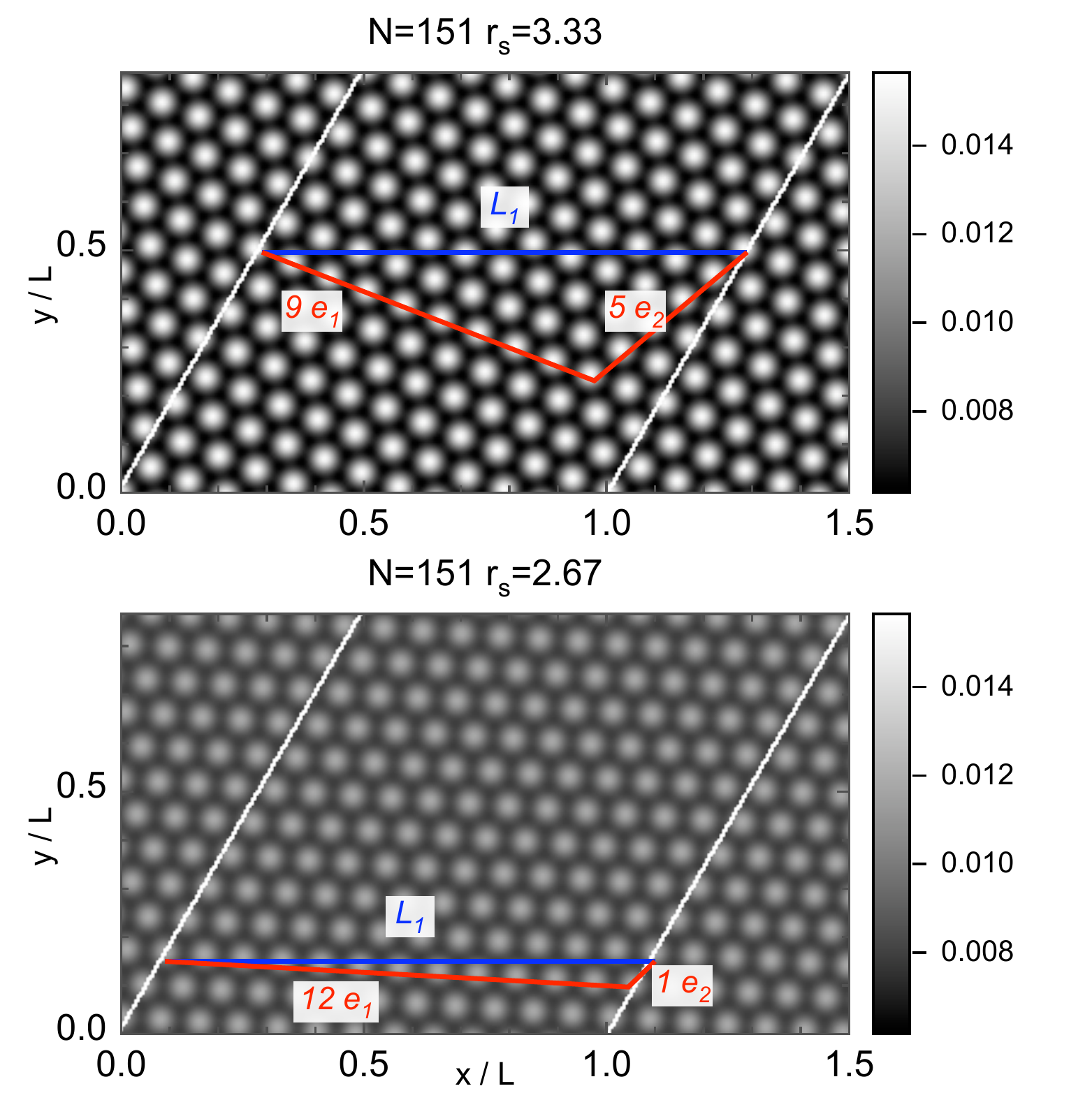}
\caption{Charge density $\rho/\langle\rho\rangle-1$, with $\rho(x)=\sum_{i=1}^N |\ffi_i(x)|^2$. Top: the number of maxima is $N_c=9^2+5^2+9\times5=151$. Bottom:  the number of maxima is $N_c=12^2+1^2+12\times1=157$. Gray levels corresponds to the same density in both figures.
Colored lines correspond to $\bL_1=l {\bf e}_1+m{\bf e}_2$, where the numbers stands for $l$, $m$ (see text).
}
\label{FIG-chargedensities}
\end{center}
\end{figure}
As $r_s$ decreases from 26.7 down to 3, only the first layer of $q$-vectors corresponding to the triangular lattice remain important. 
In this region, 
only the hexagon of the 6 lowest $q$-vectors of the WC remains (see Fig. \ref{FIG-allrs}),
and the shape of the charge-charge density in $k$-space is hexagonal instead of the circular shape
at lower densities. 
This wavefunction has therefore a similar character as the hybrid wave functions
studied in Ref.~\cite{Waintal}.
However, in contrast to Ref.~\cite{Waintal}, instead of a phase transition, we find a continuous change from many layers 
of $k$-vectors compatible with the WC at high $r_s$
to only one layer  (Figure \ref{FIG-allrs}) as $r_s$ decreases. 
At the same time, $\tau(\bq)$ in Eq.\,(\ref{EQ-PHIPERP})
is dominated by the contribution of the
FG vectors $\bk_0$ close to Fermi surface. For $|\bk_0| \approx k_F$,  $\phi_{\bk_0}^\perp(\bk)$ has essentially only one nonzero component 
at the vector $\bk=\bk_0+\bq$  closest to the Fermi surface,
with $\bq$ in the reciprocal lattice of the WC.

\begin{figure}
\begin{center}
\includegraphics[width=.49\textwidth]{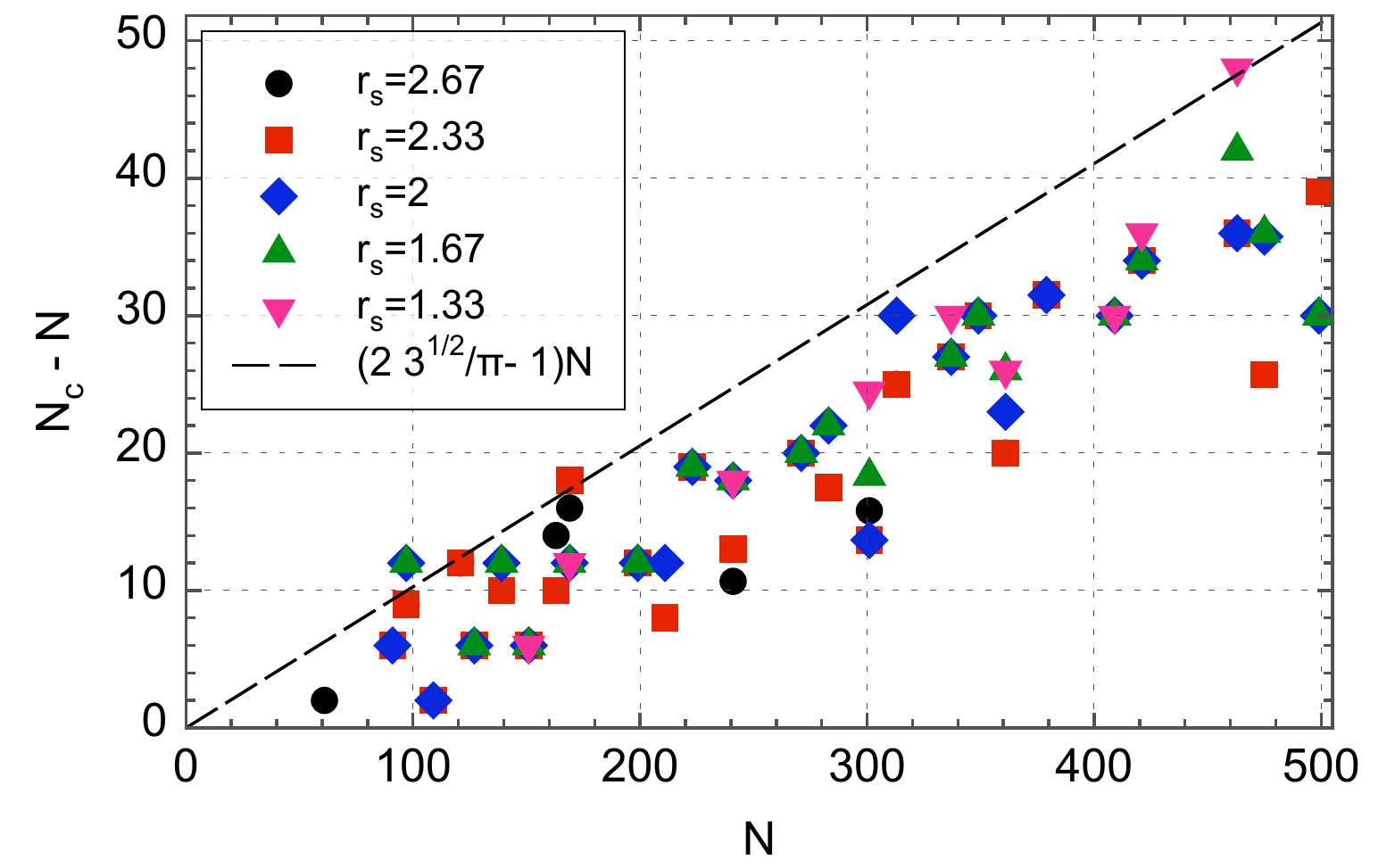}
\caption{$N_c-N$ versus the number $N$ of electrons, where $N_c$ is the number of maxima in the charge density. Remark that $N_c$ never decreases with $r_s$. The line is from Eq.\,(\ref{EQ-Nc}).
}
\label{FIG-p}
\end{center}
\end{figure}

At $r_s\approx 2.7$, the system lowers its energy by delocalizing the electrons on a denser lattice with more sites than electrons (Figure \ref{FIG-allrs} and \ref{FIG-chargedensities}). This denser lattice 
is characterized by integer numbers $(l',m')$ different from the WC lattice $(l,m)$. 
For some system sizes $N$, the maxima of $\tau(\bq)$ correspond to various couples of $\{l',m'\}$ leading to different
number of lattice sites, $N'=l'^2+m'^2+l'm'$ (Fig.\ref{FIG-p}). In this case, our system looks like a periodic crystal with an incomplete band in contrast to the WC solutions of fully occupied bands,
studied in Ref. \cite{Needs}. Therefore, our translational symmetry breaking
HF solutions below $r_s \approx 2.7$, have 
a genuine metallic character. However, as $r_s$ approaches zero, the energy gain of this metallic
crystal compared to the FG gets more and more tiny. 
At the same time, as one can see in Fig.\ref{FIG-p}, $N_c$ is either constant or increases when $r_s$ decreases (apart for one exception $N=301$ at $r_s=2.67$). For $r_s<1$ and $N<500$, the FG is always the ground state.
 
Now we show that the translational symmetry breaking, metallic states should exist at small $r_s$ in the thermodynamic limit.
Let us replace a plane wave state $\bk$ of the free FG ($\| \bk \| \le k_F$)
by a superposition of two plane waves with
wavevectors $\bk$ and $\bk+\bq$ ($\| \bk +\bq\| > k_F$). 
Choosing $\bq$ on the
six-fold star of a triangular lattice we certainly obtain a gain in potential energy.
The increase of kinetic energy is minimized if $\|\bk\|\sim k_F$ and $\| \bk +\bq\| \sim k_F$.
Then, the number of solutions for $\bk$ is optimal if $\|\bq\|\sim 2 k_F$.
This solution corresponds  to a triangular lattice of length $L_c=2 \pi/(\sqrt{3} k_F)$ in real space leading to a unit cell of volume $\Omega_c=\sqrt{3}L_c^2/2$. 
Since our system is contained in the volume $\Omega= \sqrt{3}L^2/2$,
we will obtain $N_c=\Omega_c/\Omega$ lattice sites, or
\begin{equation}
\label{EQ-Nc}
N_c = \frac{2\sqrt{3}}{\pi} N \approx 1.1 N
\end{equation}
where we have used $n=N/\Omega=k_F^2/(4\pi)$. From Fig. \ref{FIG-p} 
we see, that, as $r_s$ decreases, the number 
of sites of the HF ground state indeed approaches the limit of Eq.\,(\ref{EQ-Nc}).

A simple model accounting for such a perturbation\cite{Enerrs0}  
(actually far in norm from the FG)
gives an energy difference from the FG varying as $-\exp(-a/r_s)/r_s$ at the limit $r_s\to0$. Our data are reasonably well represented by such a law (see Fig.~\ref{FIG-Comparison}).

In conclusion,
we  have computed the ground state energy of the two dimensional fully polarized periodic electron gas
in the HF approximation.
We recover the WC at large $r_s$.
For $r_s \lesssim 2.7$, the ground state changes symmetry. Still breaking translational symmetry,  the crystal formed contains more lattice sites than particles and should therefore have metallic properties. For our finite systems, this solution disappears at small $r_s$ in favor of the FG. 
The same metallic phase has also been numerically observed as the HF ground state both in the  unpolarized system and in a model where the electrons interact with a screened Coulomb potential\cite{Enerrs0}. 

Post HF approaches, like
the Quantum Monte Carlo method\cite{Tanatar,Bernu,Waintal},
have to be used in order to study the influence of correlation effects and to establish precisely the phase diagram of the electron gas.

Acknowledgment: We thank J. Trail and R. Needs for providing us numerical data of Ref.~\cite{Needs} shown in Fig.\ref{FIG-Comparison}.


\begin{thebibliography}{99}
\bibitem{LatticeModulus} In the triangular triangular, the magnitude of the 
reciprocal lattice vectors are scaled  $\| \bK_1 \| =\| \bK_2 \| = (2/\sqrt{3})2 \pi /L $, compared to a rectangular lattice.
\end{thebibliography}
\end{document}